\documentclass[useAMS,usenatbib]{mnras}
\usepackage{graphicx,psfig,cite,epsf}  
\usepackage{times}
\usepackage{subfigure}
\usepackage{textcomp}
\usepackage{multirow}
\usepackage{amsmath}	
\usepackage{amssymb}	
\usepackage{gensymb}

\def\apjl{ApJ}                
\def\apjs{ApJS}               
\def\ao{Appl.~Opt.}           
\def\aap{A\&A}                

\def\src{XTE J1810-197}
\def\xmm{{\it XMM-Newton}}
\def\chandra{{\it Chandra}}
\def\nicer{{\it NICER}}

\def\rosat{{\it ROSAT}}

\title[The quiescence of XTE J1810-197]{The 11 years of low-activity of the magnetar XTE J1810-197}
\author[Fabio Pintore]{Fabio Pintore$^1$, Sandro Mereghetti$^1$, Paolo Esposito$^1$, Roberto Turolla$^{2,3}$,\newauthor Andrea Tiengo$^{4,1,5}$, Nanda Rea$^{6,7}$, Federico Bernardini$^{8,9,10}$, Gian Luca Israel$^8$ \\
$^1$ INAF $-$ IASF Milano, via E. Bassini 15, I-20133 Milano, Italy \\
$^2$ Dipartimento di Fisica e Astronomia, Universita di Padova, via F. Marzolo 8, I-35131 Padova, Italy \\
$^3$ Mullard Space Science Laboratory, University College London, Holmbury St. Mary, Dorking, Surrey, RH5 6NT, UK \\ 
$^4$ Scuola Universitaria Superiore IUSS Pavia, piazza della Vittoria 15, I-27100 Pavia, Italy\\
$^5$ Istituto Nazionale di Fisica Nucleare, Sezione di Pavia, via A. Bassi 6, I-27100 Pavia, Italy \\
$^6$ Instituto de Ciencias de l'Espacio (ICE, CSIC-IEEC), Carrer de Can Magrans, S/N, 08193, Barcelona, Spain \\
$^7$ Institut d'Estudis Espacials de Catalunya (IEEC), Gran Capit\`a 2-4, 08034, Barcelona, Spain \\
$^8$ INAF $-$ Osservatorio Astronomico di Roma, via Frascati 33, I-00040 Monteporzio Catone, Roma, Italy \\
$^9$ INAF $-$ Osservatorio Astronomico di Capodimonte, Salita Moiariello 16, I-80131 Napoli, Italy\\
$^{10}$ New York University Abu Dhabi, Saadiyat Island, Abu Dhabi, 129188, United Arab Emirates\\
}

\begin{document}

\maketitle

\begin{abstract}

In 2003, the magnetar \src\ started an outburst that lasted until early 2007. In the following 11 years, the source stayed in a quiescent/low activity phase. \src\ is one of the closest magnetars, hence its X-ray properties can be studied in detail even in quiescence and an extended monitoring has been carried out to study its long term timing and spectral evolution.
Here, we report the results of new X-ray observations, taken between September 2017 and April 2018, with \xmm, \chandra\ and \nicer. 
We derived a phase-connected  timing solution yielding a frequency derivative of $-9.26(6)\times10^{-14}$ Hz s$^{-1}$.  This value is consistent with that measured between 2009 and 2011, indicating that the pulsar spin-down rate remained  quite stable during the long quiescent period. A spectral analysis of all the X-ray observations taken between 2009 and 2018 does not reveal  significant spectral and/or flux variability. The spectrum of \src\ can be described by the sum of two thermal components with temperatures of 0.15 and 0.3 keV, plus a power law component with photon index 0.6.
We also found evidence for an absorption line at $\sim$1.2 keV and width of 0.1 keV.
Thanks to the long exposure time of the summed \xmm\ observations, we could also carry out a phase-resolved spectral analysis for this source in quiescence. This showed that the flux modulation can be mainly ascribed to the the warmer of the two thermal components, whose flux varies by $\sim45$ per cent along the pulse phase.

\end{abstract}


\section{Introduction}
\label{intro}

Magnetars are isolated neutron stars (NSs) with magnetic fields generally higher than $10^{14}$ G and a X-ray/soft $\gamma$-ray emission believed to be powered by the decay and instability of their extreme internal magnetic fields \citep[e.g.][]{duncan92,paczynski92,thompson01}. They have X-ray luminosities and spin periods in the ranges $L_X \sim 10^{31}-10^{36}$ erg s$^{-1}$ and $P$$\sim$0.3-12 s, respectively. Most of these sources are strongly variable showing, at  unpredictable times, large outbursts during which their X-ray flux increases up to three orders of magnitude and then decays on a variety of timescales \citep{esposito18}.
Typically, magnetars X-ray   spectra are  well described by the sum of a thermal component, believed to originate from (a region of) the NS surface, and a power law component, associated to repeated resonant scatterings of the soft thermal photons by relativistic electrons flowing in the magnetosphere. In some cases, additional spectral components are necessary \citep[see e.g.][for recent reviews]{mereghetti09,mereghetti15,kaspi17,turolla15,esposito18}.

Although  already detected by \rosat\ in 1993 as a weak source with  a 0.5--10 keV flux of $(5-10)\times10^{-13}$ erg cm$^{-2}$ s$^{-1}$, \src\ remained unnoticed until it experienced a powerful outburst discovered by RXTE in 2003 (\citealt{gotthelf04,ibrahim04}). 
Since  the initial phases of the outburst were missed, it was possible to set only a lower limit on the peak flux (a factor of 100 higher than the quiescent level). The outburst was followed with a multi-wavelength monitoring, and it was possible to discover X-ray pulsation at $\sim5.54$ s and to measure a source spin down of $6.7\times10^{-13}$ s s$^{-1}$ \citep{ibrahim04}. In addition, several short bursts were seen during the initial phases of the outburst decay \citep{woods05}. These properties indicated that  \src\ could be interpreted as a magnetar. 
 \src\ was also the first magnetar detected as a radio pulsar in the radio band \citep{hgb05}, where, during the latest phases of the decay, it showed intense and pulsed emission in phase with the X-ray pulsations (\citealt{camilo06}, \citealt{camilo07}; {even though it is likely that pulsed radio emission was present also during the initial phases of the outburst}). The outburst lasted until early 2007, when the source returned at a flux level similar to that of the pre-outburst epochs \citep[e.g.][]{alford16,pintore16}, although its radio emission continued till late 2008 \citep[e.g.][]{camilo16} {and a decaying flux from a hot region on the NS surface was present until 2009 \citep{alford16}}. 

Thanks to the continuous monitoring of \src\ carried out in the radio and X-ray bands, it was possible to investigate its  spin period variability during the outburst decay and in quiescence. The source showed a high and variable spin-down rate during the outburst decay (between $-1\times10^{-13}$ and $-5\times10^{-13}$ Hz s$^{-1}$) and a more stable spin-down rate during the quiescent phase ($\sim -9.2\times10^{-14}$ Hz s$^{-1}$; \citealt{pintore16,camilo16}). 

The spectral properties of \src\ during the outburst decay could be well modelled by the sum of three blackbody components with temperatures of $\sim0.15$, 0.3 and 0.7 keV. They were associated to the whole NS surface and to two concentric  hot-spots on the NS surface \citep{bernardini09short,albano10,alford16,pintore16,cotizelati17}. 
During the quiescent phase following the outburst, the spectrum could be fit with only two blackbody components \citep[e.g.][]{bernardini09}. Note that the quiescent spectrum seen with ROSAT before the 2003 outburst could be fit with a single blackbody, but this might be due to the limited bandwith and counting statistics of the data \citep[e.g.][]{bernardini09}.

Here we first report the spectral and timing analysis of a new set of \xmm, \chandra\ and \nicer\ observations taken between June 2017 and April 2018 and then we use the whole dataset of the long quiescent period (2007--2018) to carry  out a sensitive  spectral analysis.

\section{Data Reduction}
\label{data_reduction}

\subsection{XMM-Newton} 

\begin{table}
  \begin{center}
\footnotesize
   \caption{Log of the \xmm, \chandra\ and \nicer\ observations.} 
      \label{log}
   \begin{tabular}{l c c c  c c }
\hline 
  Obs.    &   Telescope &Obs. ID  & { Epoch$^a$} & Duration\\
No. &            &             &  MJD    & ks  \\
\hline
1 & \xmm\ & 0552800201  &  54895.6543341  &  63.6\\
2 & \xmm\ & 0605990201  &  55079.6256771  &  19.4\\
3 & \xmm\ & 0605990301  &  55081.5548494  &  17.7\\
4 & \xmm\ & 0605990401  & 55097.7062563 & 12.0\\
5 & \xmm\ & 0605990501  &  55295.1863453  &   7.7\\
6 & \xmm\ & 0605990601  &  55444.6796630  &   9.1\\
7 & \xmm\ & 0671060101  &  55654.0878884  &  17.4\\
8 & \xmm\ & 0671060201  &  55813.3872852  &  13.7\\
9 & \xmm\ & 0691070301  &  56176.9826811  &  15.7\\
10 & \xmm\ & 0691070401 &  56354.1968379  &  15.7\\
11 & \xmm\ & 0720780201 &  56540.8584298  &  21.2\\
12 & \xmm\ & 0720780301 &  56720.9705351  &  22.7\\
13 & \nicer\ & 0020420104 &  57929.3250089   & 0.8 \\
14 & \nicer\ & 0020420105 &  57930.0893007  & 0.4 \\ 
15 & \nicer\ & 0020420106$^*$ &  57932.3480273    & 2.7 \\
16 & \nicer\ & 0020420107 &  57934.9969029    &  0.2 \\
17 & \nicer\ & 0020420108$^*$ &  57938.0619294    &  2.1 \\
18 & \nicer\ & 0020420109$^*$ &  57939.3048476    &  1.0 \\
19 & \nicer\ & 0020420111$^*$ &   57947.2165493   &  3.0 \\
20 & \nicer\ & 0020420112 &  57948.1692505    &  0.7\\
21 & \nicer\ & 1020420102$^*$ &  57975.1431786    &  1.3\\
22 & \nicer\ & 1020420103 &  57976.3637636    &  0.9\\
23 & \nicer\ & 1020420104$^*$ &  57980.7382669    &  0.8\\
24 & \nicer\ & 1020420105 &  57981.0546200   &  0.5\\
25 & \nicer\ & 1020420106 &  57982.0847836    &  2.2\\
26 & \nicer\ & 1020420107 &  57983.1132332    &  1.3\\
27 & \nicer\ & 1020420108$^*$ &  57984.4005154    &  1.0\\
28 & \nicer\ & 1020420109$^*$ &  57985.3003735    &  0.8\\
29 & \nicer\ & 1020420110 &  57987.3589781    &  0.3\\
30 & \nicer\ & 1020420111 &  57988.1462045   &  0.5\\
31 & \nicer\ & 1020420112$^*$ &  57990.1375294    &  1.6\\
32 & \nicer\ & 1020420113$^*$ &  57991.3490457    &  1.8\\
33 & \nicer\ & 1020420114$^*$ &  57992.3801485    &  3.0\\
34 & \nicer\ & 1020420115$^*$ &  57993.1526852    &  1.6\\
35 & \xmm\ & 0804590201 &  58002.0363052  &  16.5\\
36 & \xmm\ & 0804590301 &  58003.0716635  &  11.2\\
37 & \xmm\ & 0804590401 &  58005.9931466  &  16.6\\
38 & \xmm\ & 0804590501 &  58011.8023528  &  10.7\\
39 & \xmm\ & 0804590601 &  58019.9415131  &  20.7\\
40  & \nicer\ & 1020420116 &  58055.2523809   &  0.6\\
41 & \nicer\ & 1020420117 &  58056.4744915    &  1.3\\
42 & \chandra\ &20091       & 58059.3379341 & 22.6\\
43  & \nicer\ & 1020420119 &  58061.0605071    &  0.6\\
44 & \nicer\ & 1020420121$^*$ &  58063.1811345    &  2.0\\
45 & \nicer\ & 1020420122&  58158.9468048    &  0.3\\
46 & \nicer\ & 1020420123 &  58163.1145907    &  0.4\\
47 & \nicer\ & 1020420124 &  58164.4848159    &  1.0\\
48 & \nicer\ & 1020420125 &  58165.1774763    &  0.7\\
49 & \nicer\ & 1020420126$^*$ &  58166.0236104    &  1.1\\
50 & \xmm\ & 0804590701 &  58180.8521837  &  10.4\\
51 & \nicer\ & 1020420127 &  58213.6570253    &  0.5\\
  \hline
\end{tabular}
\end{center}
{ $^a$ Mean time of the observation. $^*$ \nicer\ observations where source pulsation was detected.}
\end{table}

We analyzed 18 \xmm\ observations taken between March 2009 and March 2018 (see Table~\ref{log}; the 2017--2018 observations are reported here for the first time).
For each observation, we reduced the data of the EPIC-pn and the two EPIC-MOS cameras using  {\sc sas} v.16.1.0. We excluded the pixels at the CCD edges (FLAG=0), selected single- and double-pixel events (i.e. {\sc pattern}$\leq$4) and single- and multiple-pixel events (i.e. {\sc pattern}$\leq$12) for pn and MOS, respectively. We extracted source and background counts from circular regions of radii of 35$''$ and 60$''$, respectively. 
For the spectral analysis, we filtered the data excluding time intervals with high background  and we rebinned the spectra   to have at least 100 counts per bin.

\begin{figure}
\center
\includegraphics[width=6.0cm,angle=270]{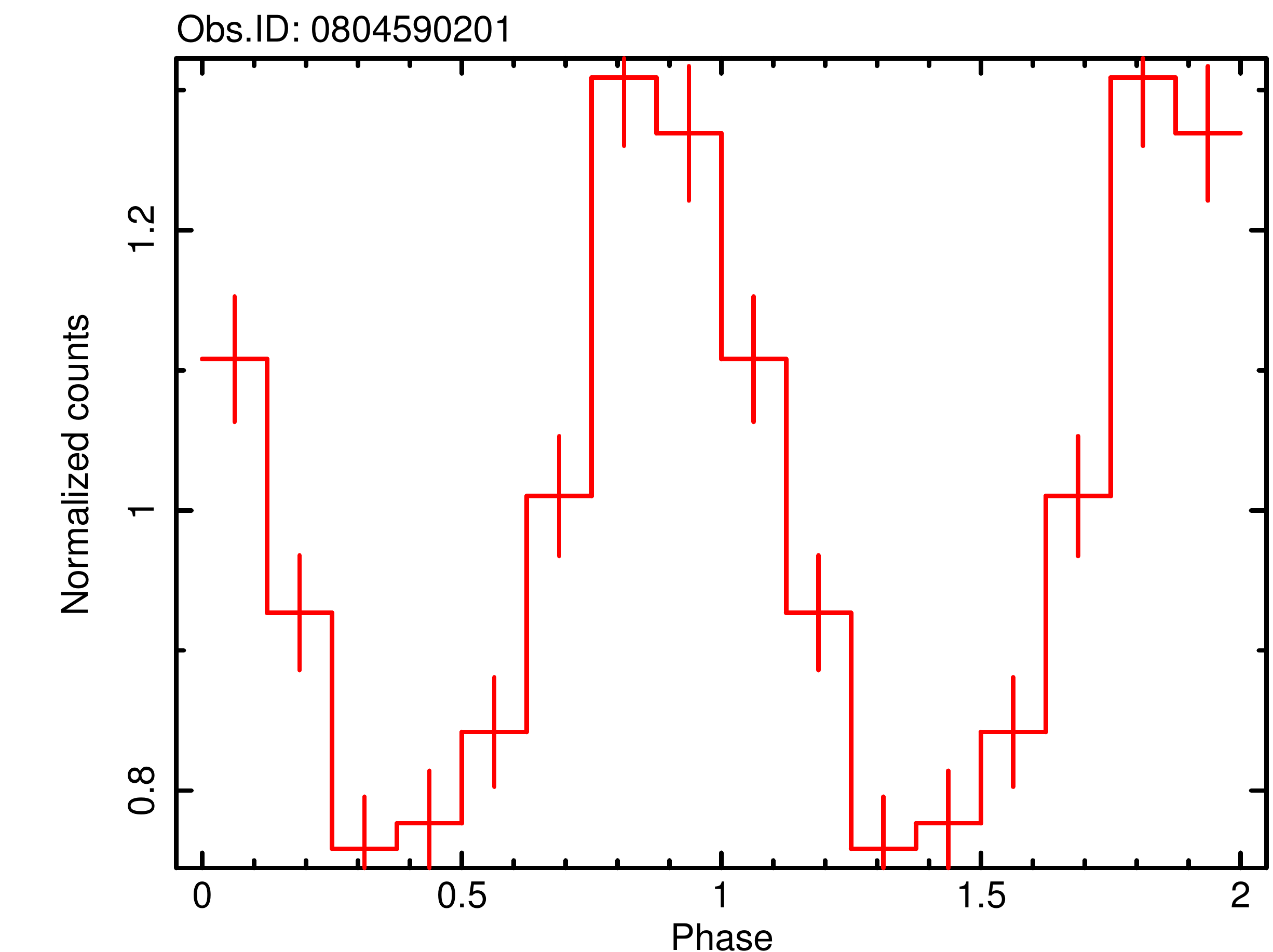}
\caption{Pulse profile of \src\ in the 1-6 keV  range obtained in the \xmm\ observation 0804590201.  The pulse profiles of all the other  2017-2018 observations are very similar. They can be well described by a single sinusoid with average pulsed fraction of $\sim30$ per cent.}
\label{pulse1}
\end{figure}

We corrected the source photons times of arrival to the solar system barycenter adopting the source most accurate coordinates R.A. = 18$^h$ 09$^m$ 51.09$^s$, Dec. = --19$^{\circ}$ 43$'$ 51.9$''$ \citep{camilo06}. No source bursts were detected during any of the observations reported here.
Comparison of the pn and MOS data showed that 
the \xmm\ observation Obs.ID=0804590401 was affected by an   instrumental problem which causes the shift of 1 second in the event times of the pn detector (the MOS is not affected by this issue, see e.g. \citealt{martin12} ). We corrected this problem by adding a second to the pn event times.

The RGS data of all observations were reduced following the standard procedure\footnote{https://www.cosmos.esa.int/web/xmm-newton/sas-thread-rgs}. For each dataset, we extracted the source spectra and grouped them with at least 30 counts per bin.

\subsection{Chandra} 

We analyzed a \chandra\ ACIS-S observation taken on 2017 November 2, with an exposure time of $\sim20$ ks (see Table~\ref{log}). We used {\sc ciao} v.4.9 and calibration files CALDB v.4.7.6. to perform the data reduction. We extracted source and background events from circular regions of radii of 3$''$ and 15$''$, respectively, and we barycentered the data with the task {\sc axbary}. Also in this case, no source bursts were found.
The source spectra were produced with the task {\sc specextract}, which generates the corresponding response and auxiliary files for the spectral analysis. Spectra were rebinned with at least 25 counts per bin.

\subsection{NICER}

We analyzed all the available \nicer\ \citep[e.g.][]{gendreau12} observations taken between June 2017 and April 2018 (see Table~\ref{log}).
We extracted the data with {\sc nicerdas} version 2018-02-22 (v2d) and adopting the tool {\sc nicerl2}. We then barycentered the data with the {\sc barycorr} task. These datasets were used only for the timing analysis because of the lack of imaging capabilities, which precludes the extraction of a properly background-subtracted spectrum for \src.

\section{Results}

\subsection{Timing}

Analysis of the source spin frequency based on \xmm\ and \chandra\ observations between 2003 and 2014 has been already reported in \citet{bernardini09}, \citet{alford16}, \citet{pintore16} and \citet{camilo16}. Therefore, we analyzed only the 2017-2018 \nicer, \chandra\ and \xmm\ observations, performing a Z$^2$ search around the expected spin frequency. We selected the energy range 1--6 keV, which yields the highest signal-to-noise ratio.
The source pulsations were significantly detected in all \xmm\ and \chandra\ observations, while only a subset of the \nicer\ observations  (shown in Table~\ref{log}) had high-enough counting statistics to allow the pulse detection. The  average spin frequency in  the whole 2017--2018 dataset is  $0.180461(1)$ Hz.  The pulse profile can be  modelled by a single sinusoidal component with average pulsed fraction\footnote{Defined as the $(A_{max}-A_{min})/(A_{max}+A_{min})$, where $A_{max}$ and $A_{min}$ are the maximum and minimum amplitude of the pulse profile, respectively.} of $28\pm2$ per cent (Figure~\ref{pulse1}).

\begin{figure}
\center
\includegraphics[width=8.7cm]{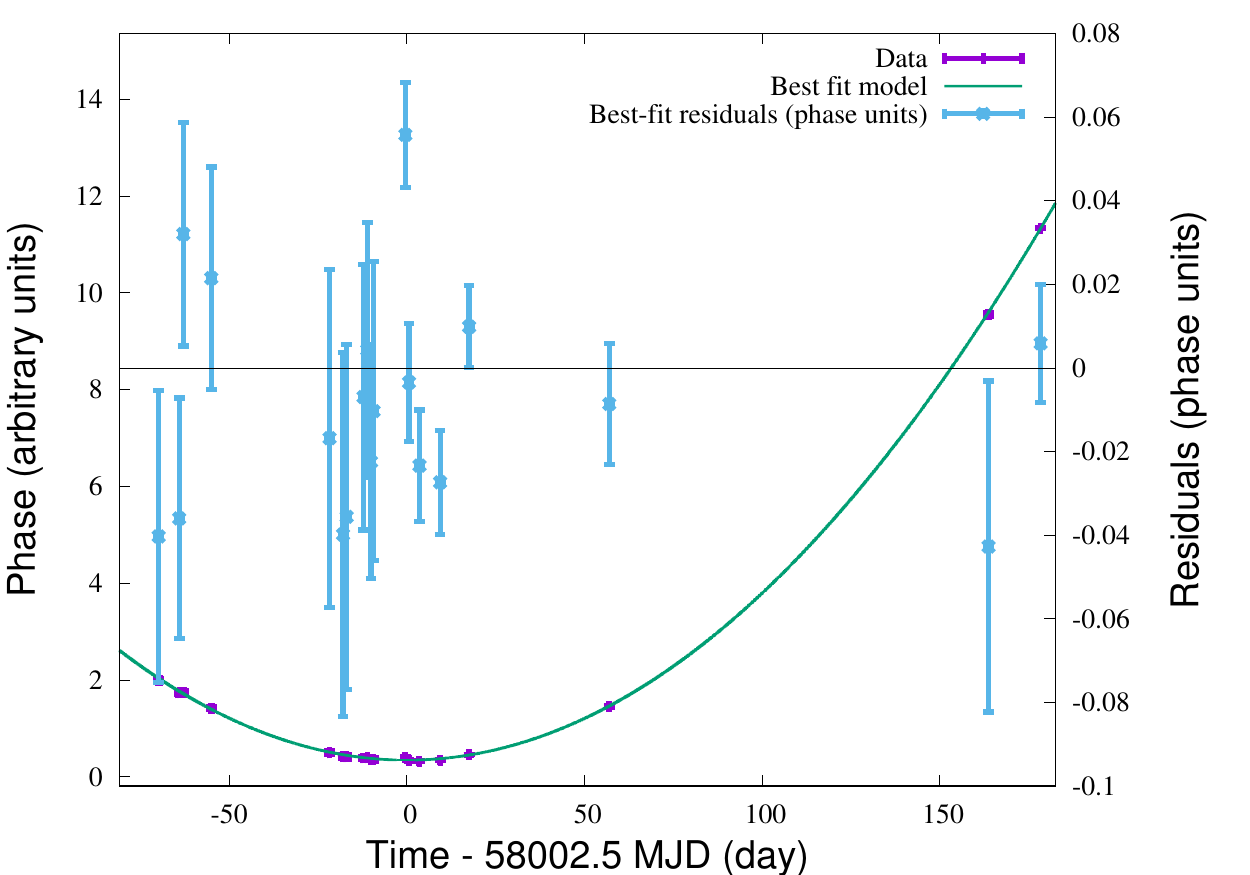}
\caption{Timing solution of the 2017-2018 \nicer, \chandra\ and \xmm\ observations (purple points). Cyan points are the residuals in phase-units of the best-fit solution (green line) with $\nu_0=0.180461427$ Hz and $\dot{\nu}=-9.26\times10^{-14}$ Hz s$^{-1}$, for T$_{0}=58002.5$ MJD. }
\label{phase_fig}
\end{figure}

To determine the spin period evolution, we initially phase-connected the pulse phases of the \nicer\ observations between 25-28 August 2017 (observations \#31--\#34), separated in time by $\sim$1d (see Table~\ref{log}). We fitted the pulse phases with a linear function of the form $\phi(t)=\phi_0+\nu_0(t-T_0)$, where $\nu_0$ is the spin frequency at the reference epoch $T_0$ (MJD 58002.5 in our analysis). Then, we added one by one all the other observations that could be phase-connected. After $\sim$30 days, a quadratic term of the form $\dot{\nu}(t-T_0)^2 /2$ started to be statistically significant.
Finally, we connected all the observations from June 2017 to April 2018, finding a timing solution with $\nu_0=0.180461427(4)$ Hz and $\dot{\nu}=-9.26(6)\times10^{-14}$ Hz s$^{-1}$ (see Figure~\ref{phase_fig} and Table~\ref{timing_sol}). A second frequency derivative was not statistically required for this dataset. 
The timing solution cannot be extended to the \xmm\ observations obtained before 2014 because of the uncertainties on the timing parameters.

\begin{table}
  \begin{center}
\footnotesize
   \caption{Best-fit timing solution of the \xmm, \nicer\ and {\it {\it Chandra}} observations. Errors are at 1$\sigma$ and estimated after adding a systematic uncertainty to the time of arrivals in order to obtain a reduced $\chi^2$ of 1.} 
      \label{timing_sol}
   \begin{tabular}{l c c }
\hline 
 Parameter & Value & Units \\
\hline
Time range &  57932--58181 & MJD                         \\
$T_{0}$ &  58002.5 & MJD                         \\
$\nu_0$ & 0.180461427(4) & Hz   \\
$\dot{\nu}$ & $-9.26(6)\times10^{-14}$ & Hz s$^{-1}$  \\
$P_0$ & 5.5413504(1) & s  \\
$\dot{P}$ & $2.84(2)\times10^{-12}$ & s s$^{-1}$  \\
\hline
$\chi_\nu^2 (dof)$ &2.37(16)& \\
  \hline
\end{tabular}
\end{center}
\end{table}

\subsection{Spectral analysis}

We performed a spectral analysis on all the \xmm\ and the 2017 \chandra\ observations, using {\sc xspec} v.12.10.0 \citep{arnaud96}, fitting the spectra in the 0.3--10 keV  energy range. Interstellar absorption was included using the {\sc tbabs} model with the  solar abundances of  \citet{wilms00}. All the errors on the spectral parameters are   at 90 per cent confidence level.

We checked that there were no significant differences in source flux or spectral shape in the 2017 and 2018 observations and that they were consistent, within the uncertainties, with those of all the previous \xmm\ observations during quiescence (i.e. observation from $\#1$ to $\#12$). Therefore, we were allowed to stack all the EPIC-pn, EPIC-MOS and RGS data into single spectra.

\begin{figure}
\center
\hspace{-0.5cm}
\includegraphics[width=6.4cm,angle=270]{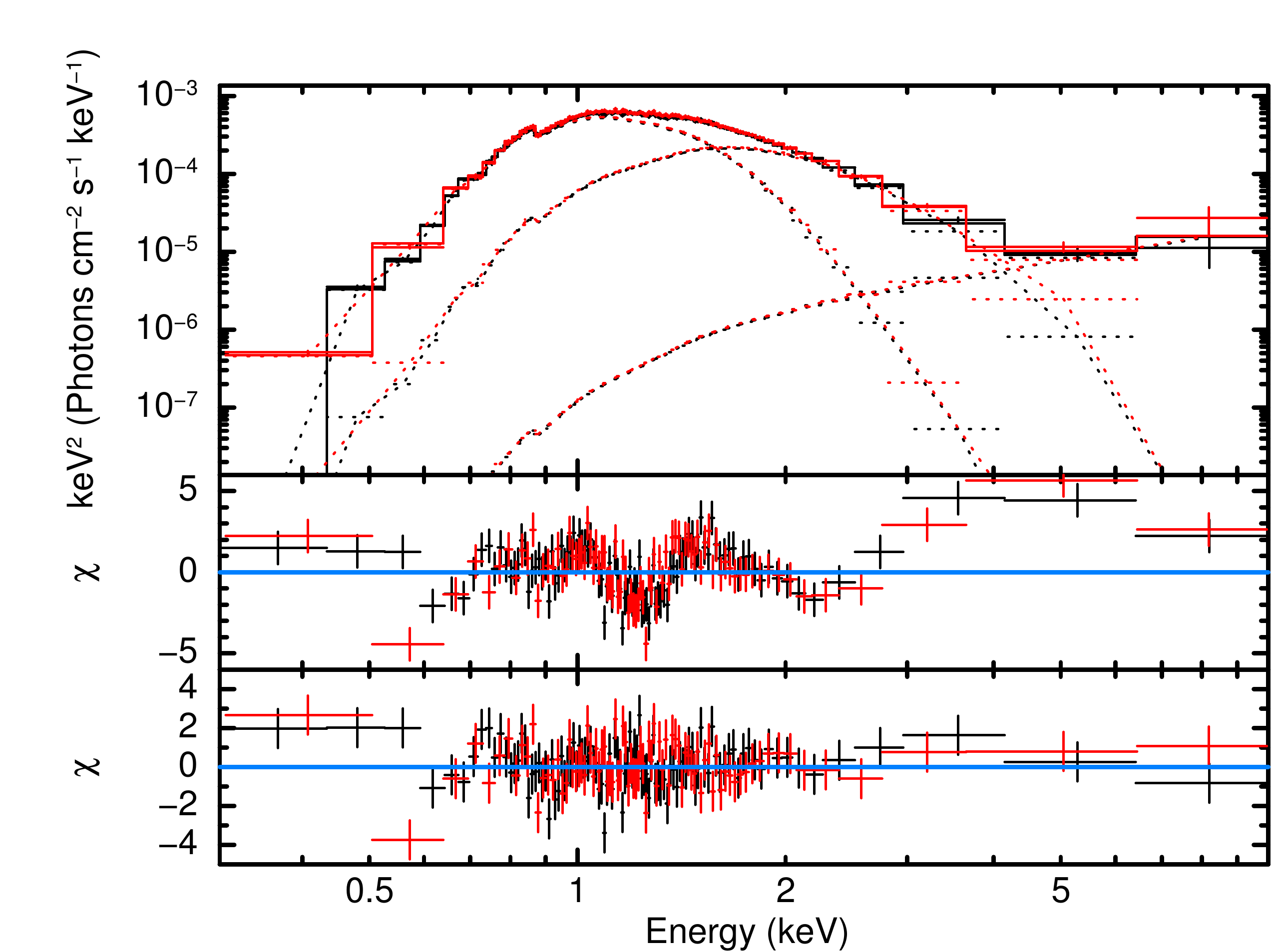}
\caption{Top panel: Stacked EPIC-pn (black) and MOS (red) spectra of   all the \xmm\ observations. 
The solid line is the best fit model ({\sc tbabs*gabs*(bbodyrad + bbodyrad + powerlaw)} in XSPEC). 
The best-fit residuals are shown in the bottom panel. 
The middle panel shows the residuals obtained without the Gaussian line and power-law components. Data have been rebinned for display purpose only.}
\label{spec_all_quiescence}
\end{figure}

The stacked EPIC-MOS and pn spectra were then fitted simultaneously.  No good fits could be obtained with either a single ($\chi^2$/dof=3646.35/879) or the sum of two blackbody components ($\chi^2$/dof=1149.86/877; Figure~\ref{spec_all_quiescence}-central panel).
We obtained an acceptable  fit ($\chi^2$/dof=897.62/872) by adding to the two-blackbody model a power law with photon index $\Gamma\sim0.6$ and an absorption line at $\sim1.2$ keV. The latter was modeled with a Gaussian profile ({\sc gabs} in {\sc xspec}). The best fit spectrum is shown in Figure~\ref{spec_all_quiescence} and all the corresponding parameters are reported in Table~\ref{spec_all_fit}. 
The  column density of $\sim1.15\times10^{22}$ cm$^{-2}$ is close to that reported in \citet{alford16} and \citet{cotizelati17}. The blackbody components have temperatures of kT$_1\sim0.15$ keV and kT$_2\sim0.3$.

We note that the power-law can be replaced with a third blackbody component, which provides a statistically acceptable fit as well ($\chi^2$/dof=897.1/872), although with poorly constrained parameters (kT = $2.5_{-0.9}^{+2.7}$ keV and emitting radius of $3_{-1.6}^{+2.9}$ meters). Such a hot blackbody was never observed in this source even during the outburst decay. In fact, the third blackbody component used to model the source spectra \citep[e.g.][]{bernardini09} had at significantly lower temperatures ($\sim0.5-0.7$ keV). For this reason, in the following we consider only the powerlaw option.

We estimated that the average absorbed 0.3--10 keV flux is $(8.04\pm0.07)\times10^{-13}$ erg cm$^{-2}$ s$^{-1}$, which is very close to the quiescent value reported in \citet{pintore16}, \citet{camilo16} and \citet{cotizelati17}, and close to the pre-outburst quiescent flux ($(5-10)\times10^{-13}$ erg cm$^{-2}$ s$^{-1}$; \citealt{gotthelf04}; \citealt{gotthelf07}). For a distance of $3.5\pm0.05$ kpc \citep{minter08}, the fluxes of the blackbody components imply emitting regions with radii of  $1.18_{-0.18}^{+0.20}$ km and $21.3_{-3.2}^{+3.4}$ km for the warmer and colder components, respectively. 
The blackbody components carry $\sim64$ and $\sim33$ per cent of the total flux for the warmer and colder component, respectively, while the powerlaw/blackbody component only $\sim3$ per cent. 

\begin{figure}
\center
\hspace{-0.5cm}
\includegraphics[width=6.1cm,angle=270]{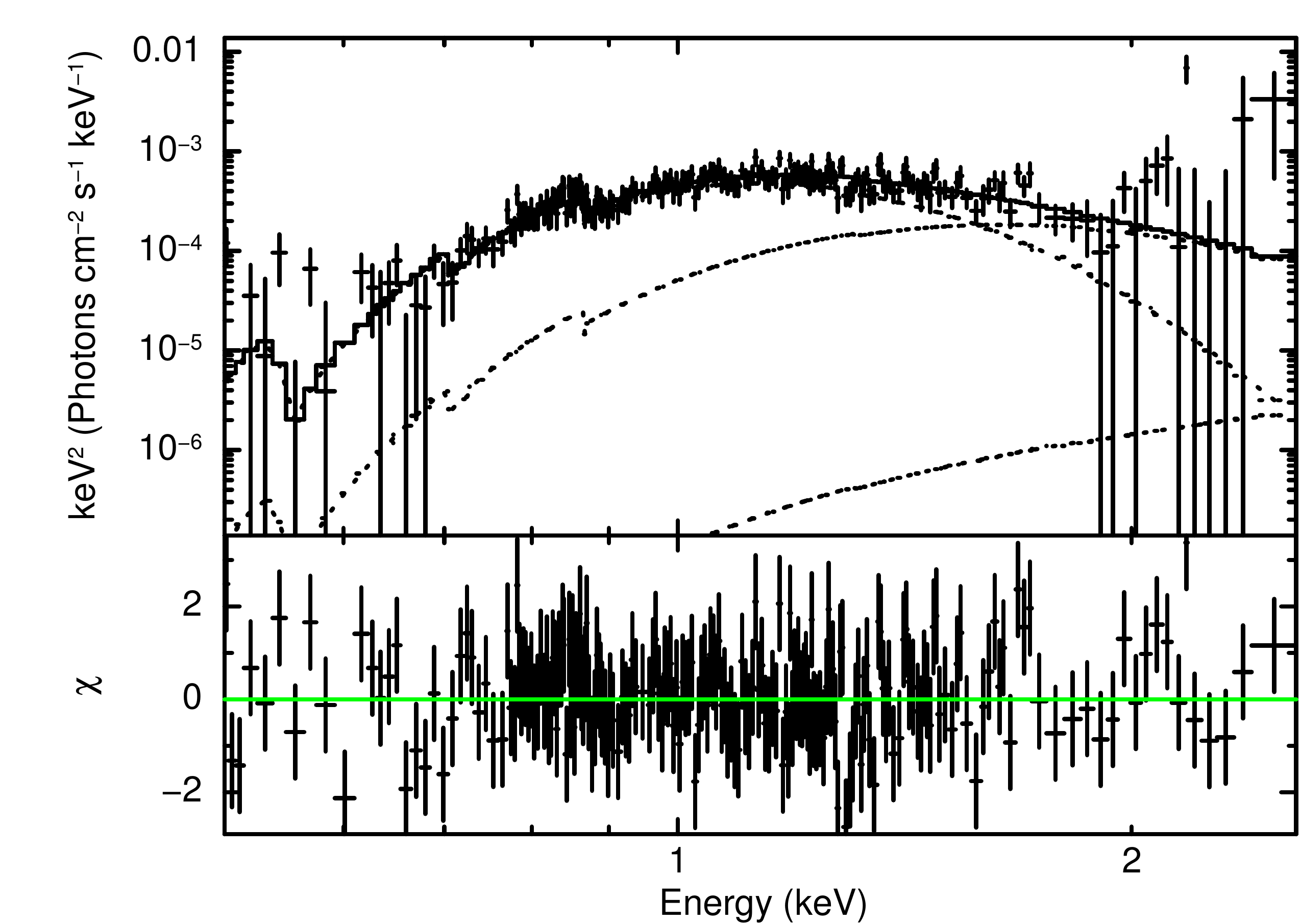}
\caption{Combined RGS spectrum of \src\  fitted with the best-fit (continuum model only) found for the EPIC-pn and MOS spectra. A weak, narrow absorption feature is seen at $\sim1.3$ keV although not statistically significant.}
\label{spec_RGS_quiescence}
\end{figure}

To check whether the line at 1.2 keV in the EPIC spectrum could be due to a blend of narrow lines, we examined the RGS spectrum. This was fitted with the same continuum model (2 blackbodies + powerlaw) used for EPIC and did not show the presence of statistically significant narrow lines around 1.2 keV (see Figure~\ref{spec_RGS_quiescence}). 

\begin{table}
  \begin{center}
  \caption{Best-fit of the stacked EPIC-pn and EPIC-MOS spectra with the {\sc tbabs*gabs(bbodyrad + bbodyrad + powerlaw)} model. Errors are at 90 per cent for each parameter of interest.}
  \label{spec_all_fit}
\begin{tabular}{lll}
\hline
Model & Component &\multicolumn{1}{c}{} \\
{\sc TBabs} & nH ($10^{22}$) & $1.16^{+0.03}_{-0.03}$  \\
{\sc bbodyrad} & kT$_1$ (keV)  & $0.143^{+0.004}_{-0.004}$ \\
 & Norm. ($10^3$) & $3.7^{+1.0}_{-0.7}$ \\
{\sc bbodyrad} & kT$_2$ (keV) &  $0.30^{+0.01}_{-0.01}$ \\
 & Norm. & $11.4^{+4.0}_{-2.8}$ \\
{\sc powerlaw} & $\Gamma$ & $0.6^{+1.1}_{-0.9}$ \\
 & Norm. (10$^{-7}$) & $8.9^{+43}_{-7.0}$ \\
{\sc gabs} & Energy (keV) & $1.24^{+0.01}_{-0.01}$ \\
 & $\sigma$ (keV) & $0.1^{+0.02}_{-0.02}$ \\
 & Strength (keV) & $0.035^{+0.01}_{-0.008}$ \\
 \hline
  $\chi^2_{\nu} (dof)$ & & 1.03(872) \\
 \hline
\end{tabular}
\end{center}
\end{table}

\subsubsection{Phase-resolved spectroscopy}

\begin{figure}
\center
\includegraphics[width=8.5cm]{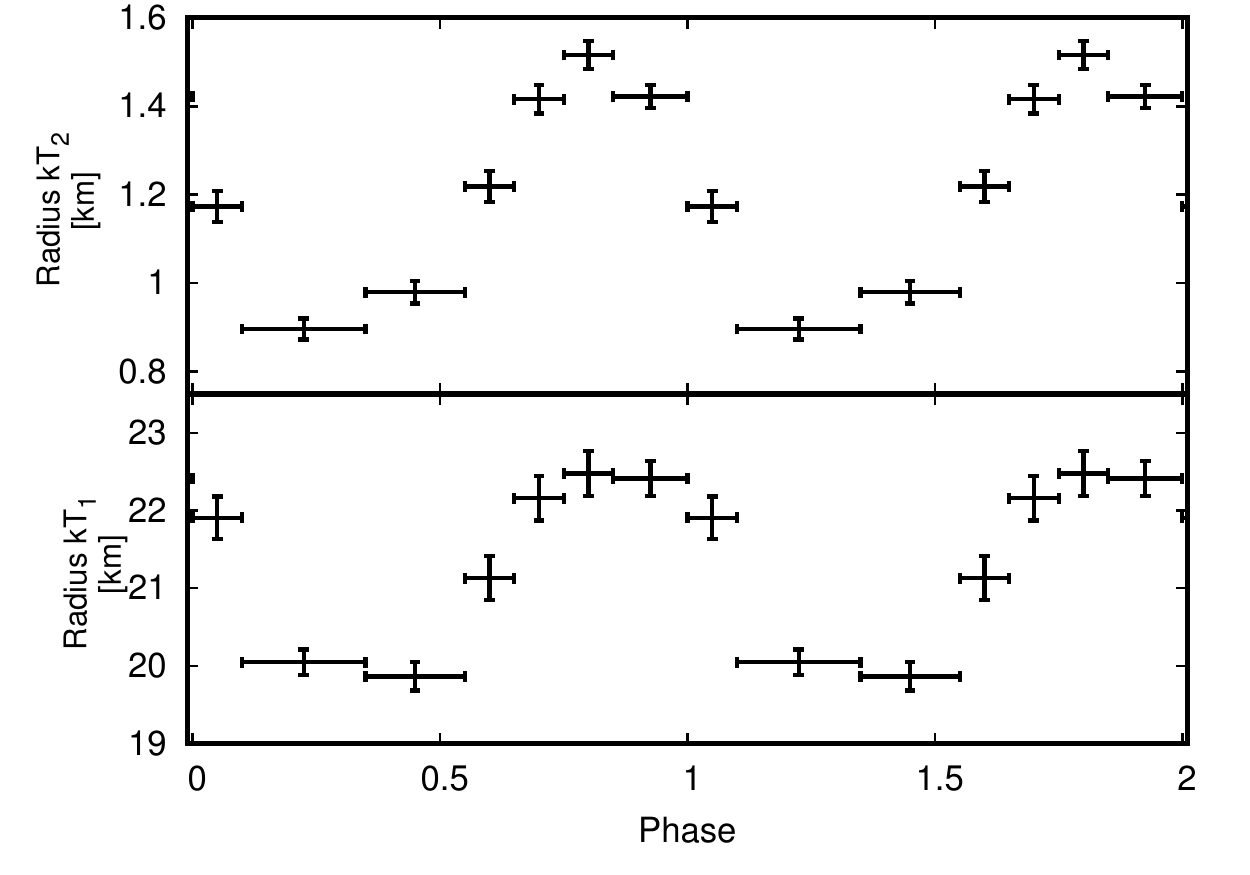}
\vspace{-0.5cm}
\caption{Emitting radius of the warmer (top) and colder (bottom) blackbodies along the pulse profile. Errors are at 90 per cent.}
\label{norm_all}
\end{figure}

We extracted  EPIC-pn   spectra for seven phase bins (see Figure~\ref{norm_all}) and fitted them simultaneously using the best-fit model of the phase-averaged spectrum with parameters fixed to those of Table~\ref{spec_all_fit}, with the addition of a multiplicative factor to  account for the different flux in each phase bin. This is clearly a poor reproduction of the spectra ($\chi^2$/dof=1370.15/197), indicating the presence of spectral variability along the pulse-profile. Therefore, we removed the multiplicative constant and we let  the normalizations of the blackbody components free to vary independently: in this way we could properly fit the data ($\chi^2$/dof=221.51/189). The normalizations  of the two blackbodies followed very well the shape of the pulse profile. The variability is larger for the warmer blackbody, for which the  normalization varies by  $\sim45$ per cent, compared to  $\sim10$ per cent for the colder one. In Figure~\ref{norm_all}, we present the emitting radius of the two blackbodies as a function of the spin-phase, for a distance of 3.5 kpc, showing a pulsed fraction of $\sim26$ and $\sim6$ per cent.
We also tried to let free to vary independently (one parameter at the time), the blackbody temperatures, the powerlaw normalization and the line normalization. The fit was generally poorly sensitive to these parameters, but we found some hint for an anti-correlation between the line intensity and the total flux. To test further the line behaviour, we extracted only two EPIC-pn spectra for the phase bins 0.75--1.2 (the minimum of the pulse profile) and 0.2--0.75 (the pulse peak) and we fitted them with the average best-fit model letting free to vary only the blackbodies and line normalization ($\chi^2$/dof=221.54/185). This analysis indicated that the line optical depth changes from $0.050\pm0.007$ at the pulse peak to $0.032\pm0.005$ at the pulse minimum, implying a measured variability at the $\sim$2$\sigma$ level. 

\section{Discussion}

\src\ was one of the first transient magnetar to be discovered  and it is the one for which it has been possible to observe the longest quiescent period following an outburst ($\sim11$ years). In fact, other transient magnetars, such as, e.g., 1E 1547.0--5408  and SGR 1627--41, displayed shorter quiescent periods, interrupted by the occurrence of recurrent outbursts \citep[see e.g.][]{cotizelati17}. In addition, since \src\ is  relatively close (3.5 kpc, \citealt{minter08}), its quiescent luminosity of  $\sim10^{33}$ erg s$^{-1}$ yields a flux sufficiently high  to permit sensitive spectral and timing studies.
 
After the decay of its outburst in early 2007, \src\ entered a low-activity phase during which the source pulsation could be still significantly detected. The quality of the timing data during this phase was good enough to measure precisely the spin-down  ($\dot{\nu}\sim-9.2\times10^{-14}$ Hz s$^{-1}$) and to find evidence of a second derivative term ($\ddot{\nu}\sim5.7\times10^{-23}$ Hz s$^{-2}$). This phase-connected timing solution was found to be valid for a baseline of $\sim1000$d (between 2009 and 2011), but strong timing noise made it impossible to extend it to earlier or later epochs \citep[e.g.][]{pintore16,camilo16}. Assuming that this timing solution remained valid until the time of the observations reported here, we would expect a spin frequency of 0.18046226(15) Hz in  the first 2017 \xmm\ observation. This is indeed quite close to the average frequency measured in the 2017--2018 monitoring (0.180461(1) Hz). Thanks to the new \xmm, \chandra\ observations and the dense \nicer\ monitoring, we could derive a new phase-connected timing solution characterized by a source spin-down of $-9.26(6)\times10^{-14}$ Hz s$^{-1}$, which is totally consistent within uncertainties with that reported for the years 2009-2011. No significant second derivative component was observed {and we derived 2$\sigma$ limits of $-2\times10^{-22}$  Hz s$^{-2}$ $<\ddot{\nu}<1\times10^{-21}$ Hz s$^{-2}$, which are consistent with previous estimates \citep[e.g.][]{camilo16}}. Further X-ray observations in 2019 could allow us to obtain a more precise   timing solution, which could also be extended backwards in time. We note that our timing solution does not exhibit strong timing noise during the $\sim8$ months of observations.

The spectral analysis of the new \xmm\ and \chandra\ datasets shows that \src\ spectrum did not change significantly during its long quiescent phase. It comprises two thermal components with temperatures of $\sim$0.15 keV and 0.3 keV, with associated emission radii of $\sim$ 21 km and 1.1 km (assuming a distance of 3.5 kpc) plus a powerlaw. 
Such a spectral decomposition for \src\ was adopted in \citet{gotthelf04}, \citet{bernardini09}, \citet{alford16}, \citet{cotizelati17}.
While the warmer component has an emitting radius consistent with a hot, localized spot on the NS surface, the large radius associated to the cooler blackbody is of the order of the whole NS size, or even larger. {This  could be due to the uncertainty in the distance or to the fact that we used simple blackbody models. We note that the currently available neutron star atmosphere models generally yield higher temperature and smaller radii than blackbody fits.
It  should therefore be explored if more physical spectral models, adequate for the magnetars, could give more realistic values for the emitting radius \citep[e.g.][]{zavlin04,potekhin14}.}
If the warmer component originated by a localized heating of the surface layers during the outburst, because of either Ohmic dissipation of back flowing currents in a twisted magnetosphere \citep[][]{beloborodov09} or energy release in the crust \citep{pons12}, a substantial decrease of the temperature is to be expected in a timescale of $\sim$1 year. However, in our results, the warmer component appears to be quite stable over the last 11 years and, if this is indeed a hot-spot onto the NS surface, a continuous injection of energy, possibly coming for the star interior, seems to be required \citep[e.g.][]{kaminker14,akgun18}. 
We also found evidence for a hard component that we modelled either with a powerlaw or a blackbody, which is yet not robustly constrained in both cases. We note that hints of an excess at high energy was already seen in the \xmm\ spectra presented in \citet{bernardini09}. We exclude the possibility that the component has thermal origin, as it would be too hot, associated to a tiny region on the surface and, especially, it was never observed in the past.
On the other hand, the powerlaw model is more reliable and such a component would be associated to resonant scattering of thermal photons from particles flowing in the star magnetosphere (\citealt{thompson02b}; see also \citealt{turolla15} for a review). 
Further investigations are required to constrain better the nature of such component.

Taking advantage of the large amount of X-ray data and the apparent absence of spectral variability, we also performed for the first time a phase-resolved spectral analysis of the quiescent epochs monitored with \xmm. We found that the radius of the emitting region of the warmer component varies as a function on the pulse-phase with an amplitude of $\sim26$ per cent (between 0.8--1.5 km), while spectral fits do not indicate variations of the temperature with phase. This finding corroborates  the association of this component with  a hot-spot, seen with a changing apparent emitting radius caused by the pulsar rotation. However, we  found that also the lower temperature component exhibits a change in the emitting radius, although less prominent ($\sim6$ per cent). 
Under the assumption that thermal photons come from the star surface, the fractional variation of the emitting areas depends on their size and location, as well as on the    angle   $\xi$ that the  line-of-sight makes  with the star rotation axis. Following the approach of \citet{turolla13a}, we calculated the relative change of the visible emitting areas over a rotational period using a simple emission model in which the warmer blackbody is emitted by a circular hot-spot  with aperture 5{\textdegree} at colatitude $\chi$ (the angle between the rotation axis and the magnetic field axis) and the colder component by a larger, concentric corona. The choice of considering concentric regions is motivated by the fact that the two pulsed components are aligned in phase.  Each region is assumed to be at constant temperature. The computation includes general-relativistic effects ($M=1.4M_{\odot}$, $R=10$ km) and was performed for several values of $\chi$ and $\xi$ in the range $[0,\,\pi/2]$. The main conclusion is that there are indeed geometries for which the observed values of the fractional variation are recovered, but this occurs only  if the colder region  extends over a very large fraction of the star surface.  A possible configuration reproducing  the observed pulsed fractions is obtained for  $\chi\sim 75${\textdegree}, $\xi\sim 15${\textdegree} and an  aperture of 115{\textdegree} for the colder region. This is of course an oversimplified model. It is likely that the colder blackbody component actually originates from the whole NS  with a non-uniform temperature distribution across the surface, that cannot be resolved into more  than one thermal component due to the limited sensitivity of the current data.

We also found evidence in the phase-averaged spectrum of an absorption line centered at $\sim1.25$ keV and with a width of  $\sim0.1$ keV. This feature was already reported in \citet{bernardini09}, \citet{alford16}, \citet{cotizelati17}, and possibly with an asymmetrical shape \citep{vurgun18}, and  tentatively associated to a resonant cyclotron scattering absorption line. 
If the line is due to cyclotron scattering/absorption by electrons, the implied magnetic field is 
$B=10^{12}(E_c/11.6\, {\mathrm keV}(1+z))\ \mathrm G$ which, assuming $z=0.8$, yields in the present case $B\sim 2\times 10^{11}\ \mathrm G$. This is much below the value of the B-field estimated from spin-down. On the other hand, assuming that the line is due to proton cyclotron gives a value $m_p/m_e$ times higher, $B\sim 3.5\times 10^{14}\, \mathrm G$ quite close to that inferred from timing, $B_p\sim 2.6\times 10^{14}\, \mathrm G$ \citep[e.g.][]{camilo16}. 

The phase-resolved spectral analysis indicates that the line optical depth may show an anti-correlation with the pulse-profile, the optical depth being lower at the pulse peak and larger near the pulse minimum. As no interstellar absorption line would behave in such a way, this is a further robust support to the intrinsic source origin of such a feature. In addition, this result prompts us to suggest that the line is formed in a region located above the NS surface but that is somehow displaced from the region where the pulse is produced (otherwise the optical depth of the line during the peak should show a maximum).

\section*{Acknowledgements} 

We used observations obtained with \xmm\, an ESA science mission with instruments and contributions directly funded by ESA Member States and NASA, the NASA \chandra\ and \nicer\ missions.
We acknowledge financial contribution from the agreement ASI-INAF I/037/12/0.
We acknowledges support from the HERMES Project, financed by the Italian Space Agency (ASI) Agreement n. 2016/13 U.O. FB is founded by the European Union's Horizon 2020 research and innovation programme under the Marie Sklodowska-Curie grant agreement n. 664931. P.E. acknowledges funding in the framework of the project ULTraS, ASI--INAF contract N.\,2017-14-H.0. 

\noindent We also thank the referee for his useful suggestions.

\addcontentsline{toc}{section}{Bibliography}
\bibliographystyle{mn2e}
\bibliography{biblio}

\bsp
\label{lastpage}
\end{document}